\documentclass[12pt]{article}
\usepackage{float} 
\usepackage{hyperref}
\usepackage{siunitx}
\DeclareSIUnit\angstrom{\text{\AA}}
\usepackage{caption}
\usepackage{newfloat} 
\usepackage{xcolor}

\usepackage{newtxtext,newtxmath}
\usepackage{mathtools}
\DeclarePairedDelimiter\abs{\lvert}{\rvert}%
\DeclarePairedDelimiter\ave{\langle}{\rangle}%
\usepackage{upgreek}
\usepackage{graphicx}

\usepackage[letterpaper,margin=1in]{geometry}

\renewenvironment{abstract}
	{\quotation}
	{\endquotation}

\date{}

\makeatletter
\renewcommand{\fnum@figure}{\textbf{Figure \thefigure}}
\renewcommand{\fnum@table}{\textbf{Table \thetable}}
\makeatother

\usepackage{scicite}

\usepackage{url}

\def\scititle{Model Accuracy and Data Heterogeneity Shape Uncertainty Quantification in Machine Learning Interatomic Potentials}
\title{\bfseries \boldmath \scititle}

\author{%
    Fei Shuang$^{1\ast}$,
    Zixiong Wei$^{1}$, 
    Kai Liu$^{1}$,\\
    Wei Gao$^{2,3}$,
    Poulumi Dey$^{1\ast}$\and
\begin{tabular}{@{}p{\textwidth}@{}}
    \raggedright 
    \small $^{1}$Department of Materials Science and Engineering, Faculty of Mechanical Engineering, Delft University of Technology, Mekelweg 2, Delft, 2628 CD, The Netherlands.\\
    \small $^{2}$J. Mike Walker’66 Department of Mechanical Engineering, Texas A\&M University, College Station, TX 77843, United States.\\
    \small $^{3}$Department of Materials Science \& Engineering, Texas A\&M University, College Station, TX 77843, United States.\\
    \small$^\ast$Corresponding authors. Emails: P.dey@tudelft.nl; F.Shuang@tudelft.nl
\end{tabular}
}

\begin{document} 

\maketitle


\newpage


\section*{Abstract}
\begin{abstract} \bfseries \boldmath

Machine learning interatomic potentials (MLIPs) enable accurate atomistic modelling, but reliable uncertainty quantification (UQ) remains elusive. In this study, we investigate two UQ strategies, ensemble learning and D‑optimality, within the atomic cluster expansion framework. It is revealed that higher model accuracy strengthens the correlation between predicted uncertainties and actual errors and improves novelty detection, with D‑optimality yielding more conservative estimates. Both methods deliver well calibrated uncertainties on homogeneous training sets, yet they underpredict errors and exhibit reduced novelty sensitivity on heterogeneous datasets. To address this limitation, we introduce clustering‑enhanced local D‑optimality, which partitions configuration space into clusters during training and applies D‑optimality within each cluster. This approach substantially improves the detection of novel atomic environments in heterogeneous datasets. Our findings clarify the roles of model fidelity and data heterogeneity in UQ performance and provide a practical route to robust active learning and adaptive sampling strategies for MLIP development.

\end{abstract}

\newpage

\noindent
\section*{INTRODUCTION}

Machine learning interatomic potentials (MLIPs) have reshaped computational materials science by bridging the accuracy of quantum-mechanical methods with the scale of classical molecular dynamics (MD)\cite{Friederich2021,ceriotti2022}. By learning the mapping from local atomic environments to potential energy surfaces using first-principles data, MLIPs routinely approach near-quantum fidelity at a fraction of the computational cost\cite{Zuo2020,jacobs2025}. This advance has enabled simulations of complex, previously inaccessible phenomena, from phase transformations and defect kinetics to catalyst discovery and non-equilibrium transport, at time and length scales far beyond \textit{ab initio} molecular dynamics\cite{Qamar2023,liang2023,Erhard2024}.

Unlike traditional, physically motivated functional forms such as the embedded-atom model, MLIPs are constrained by their training distributions. When presented with out-of-distribution (OOD) atomic environments, they may yield unreliable or unphysical predictions, limiting transferability in practical workflows. This challenge has motivated a rich set of uncertainty quantification (UQ) strategies to assess reliability of energies and forces. Among these, D-optimality and ensemble-based methods have been particularly influential owing to their practical implementation across multiple frameworks. The D-optimality criterion, implemented in Moment Tensor Potentials (MTP)\cite{Shapeev2016,Novikov2021,Podryabinkin2023}, the Atomic Cluster Expansion (ACE)\cite{Drautz2019,Lysogorskiy2021,Bochkarev2022}, and Neuroevolution Potentials (NEP)\cite{GPUMD_4.0}, identifies informative configurations via their contribution to feature-space volume (e.g., extrapolation grade). In parallel, ensemble approaches estimate epistemic uncertainty by measuring the spread of predictions across models trained with different initializations, data bootstraps, or hyperparameters.

Beyond their role in diagnosing reliability, UQ methods have become central to data generation via active learning. In UQ-guided loops, candidate configurations discovered during exploration are selectively labeled and appended to the training set, yielding automated, recursive improvement in both accuracy and robustness. This paradigm has matured into a standard practice for MLIP development: it reduces the size (and cost) of reference datasets while enhancing stability under demanding conditions. In applications, D-optimality-based selection within MTPs is a mainstay for metals and alloys\cite{carral2023stability, xu2024origin, rybin2025accelerating, klimanova2025accelerating, kotykhov2025actively}, whereas ensemble-force criteria are particularly effective in complex, heterogeneous systems such as silicon--oxygen networks\cite{Erhard2024}. Recent hyperactive learning strategies further accelerate sampling by biasing dynamics toward uncertain regions, efficiently generating information-rich configurations for linear ACE potentials\cite{van2023hyperactive}. Collectively, these developments underscore the pivotal role of UQ in both the application and advancement of MLIPs\cite{perez2025uncertainty, bilbrey2025uncertainty}.

Despite this progress, key questions remain regarding calibration and transferability of UQ metrics. Notably, Lysogorskiy \textit{et al.} reported within the ACE framework that D-optimality and ensemble indicators offer broadly comparable reliability\cite{Lysogorskiy2023}. Two issues are particularly pressing. First, how does the baseline predictive accuracy of a fitted MLIP influence the fidelity of its uncertainty estimates? Second, how does increasing dataset heterogeneity (e.g., mixing simple elastic deformations with defect-rich clusters, surface reconstructions, liquid-like motifs, and high-strain-rate configurations) affect the calibration and sensitivity of UQ measures? These questions are especially relevant for on-the-fly active learning, wherein the training set evolves to include progressively more diverse atomic environments, potentially improving coverage while challenging model generalization.

In this work, we systematically evaluate ensemble-based and D-optimality UQ within the ACE framework. We quantify how model accuracy and dataset heterogeneity together govern (i) the alignment between predicted uncertainties and realized errors and (ii) each method’s capability to flag novel configurations and local atomic environments (LAEs). Building on these insights, we introduce a \emph{clustering-enhanced local D-optimality} criterion: configuration space is partitioned into clusters of similar atomic motifs, and extrapolation grades are computed within each cluster rather than globally. This strategy improves calibration, tracks true errors more faithfully, and more reliably detects OOD LAEs in large-scale deformation simulations. The resulting protocol maintains the computational efficiency of ACE models while providing uncertainty estimates that are both sensitive and robust across heterogeneous datasets.

\section*{RESULTS}
\subsection*{Dataset preparation and analysis}

\begin{figure}[!ht]
    \centering
    \includegraphics[width=1\linewidth]{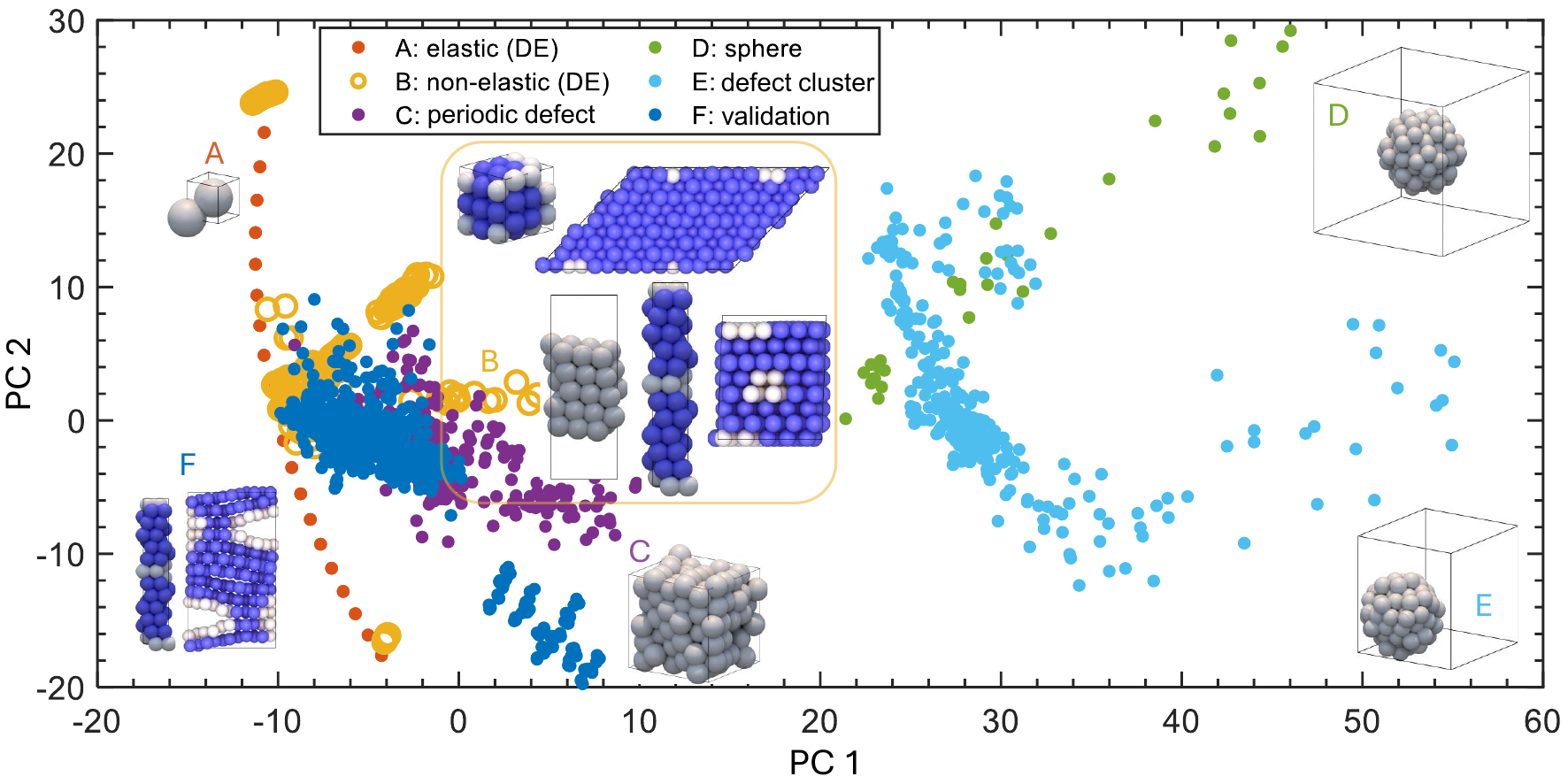}
\caption{\textbf{Six subsets of W configurations used in this study.} 
Each point represents a configuration projected onto the first two principal components of its MACE descriptor.  
\textbf{(A)} Elastic deformations;  
\textbf{(B)} Supercell configurations from AIMD simulations and defective structures (grain boundaries, vacancies, and dislocations);  
\textbf{(C)} Defect-related structures with reconstructed periodic boundary conditions;  
\textbf{(D)} BCC spherical clusters with free surfaces;  
\textbf{(E)} Atomic clusters from defective regions;  
\textbf{(F)} Test set configurations.  
Representative atomic structures for each subset are shown in the insets. The union of datasets \textbf{A} and \textbf{B} forms the domain expertise (DE) set, with \textbf{B} termed the non-elastic subset. Datasets \textbf{C}, \textbf{E}, and \textbf{F} are from Ref.~\cite{shuang2025modeling}.}
    \label{fig1-pca}
\end{figure}

We employ the body-centered cubic tungsten (BCC W) dataset from our recent work~\cite{shuang2025modeling} for UQ. Fig.~\ref{fig1-pca} displays all the configurations of the six subsets (A to F) using the first two principal components of the MACE descriptor~\cite{batatia2022mace}, with each subset annotated by its representative configuration. 
Details of these subsets are summarized below:
\begin{itemize}
    \item[\textbf{A}] Unit cells undergoing elastic deformation (two atoms per cell).
    \item[\textbf{B}] \textit{Ab initio} molecular dynamics (AIMD) snapshots and simple defects, including vacancies, dislocations, grain boundaries (GBs), and surfaces.
    \item[\textbf{C}] Atomic clusters extracted from complex defects in large-scale MD simulations, with periodic boundary conditions reconstructed using an empirical interatomic potential--guided grand-canonical Monte Carlo (EIP-GCMC) method. Methodological details are provided in \cite{shuang2025modeling}.
    \item[\textbf{D}] Spherical BCC clusters embedded in vacuum within a periodic box, introducing a large fraction of free surfaces.
    \item[\textbf{E}] Atomic clusters cut from complex defects using the MLIP-3 package \cite{Podryabinkin2023}.
    \item[\textbf{F}] A comprehensive validation set from our previous study~\cite{shuang2025modeling}, spanning diverse defect and deformation scenarios, including GBs with random perturbations, GBs under severe compression, two- and three-dimensional random GBs, and crack tip originally from Ref.~\cite{Lei-crack-npj}.
\end{itemize}
Subsets A and B together form the typical foundation for initial MLIP training through domain expertise (DE). This progression, which starts from simple elastic strains in A, moves through increasingly complex defect structures and surfaces in B to E, and culminates in the broad validation collection in F, enables systematic assessment of MLIP performance and UQ behavior across increasing configurational complexity.

In the following sections, we perform UQ analysis using two dataset combinations. The first employs A+B for training and C+E+F for testing, representing a typical scenario where MLIPs predict atomic environments for unseen defects from standard DE datasets. The second, more challenging combination uses A+D for training and B+C+E+F for testing, where elastic deformations (A) and free surfaces (D) create highly heterogeneous features. In this case, all test configurations become out-of-distribution (OOD) relative to the training set. Our results demonstrate that while both ensemble learning and D-optimality provide satisfactory UQ performance in the first scenario, they struggle with the increased complexity of the second case.

\subsection*{Ensemble learning method}

In this section, we employ the maximum deviation of ACE predictions to quantify uncertainty via the ensemble learning method, following Ref.~\cite{Lysogorskiy2023} and as detailed in \nameref{methods}. At the configurational level, we consider the configuration-based energy (CBE) and configuration-based force (CBF) criteria, quantified by \(U_{E,\mathrm{cfg}}\) and \(U_{F,\mathrm{cfg}}\), respectively. At the atomic level, we adopt the atom-based force (ABF) criterion, denoted by \(U_{F,\mathrm{atom}}\). CBE and CBF facilitate active learning or sampling of entire configurations, whereas ABF is tailored to select LAEs in large‑scale simulations. We then compute the corresponding errors \(e_{E,\mathrm{cfg}}\), \(e_{F,\mathrm{cfg}}\), and \(e_{F,\mathrm{atom}}\) (defined in~\nameref{methods}) and examine the correlation between each uncertainty metric and its error. We consider A + B as the training set with C + E + F for testing. A six‑member ensemble is employed to quantify predictive uncertainty.

\begin{figure}[!ht]
    \centering
    \includegraphics[width=1\linewidth]{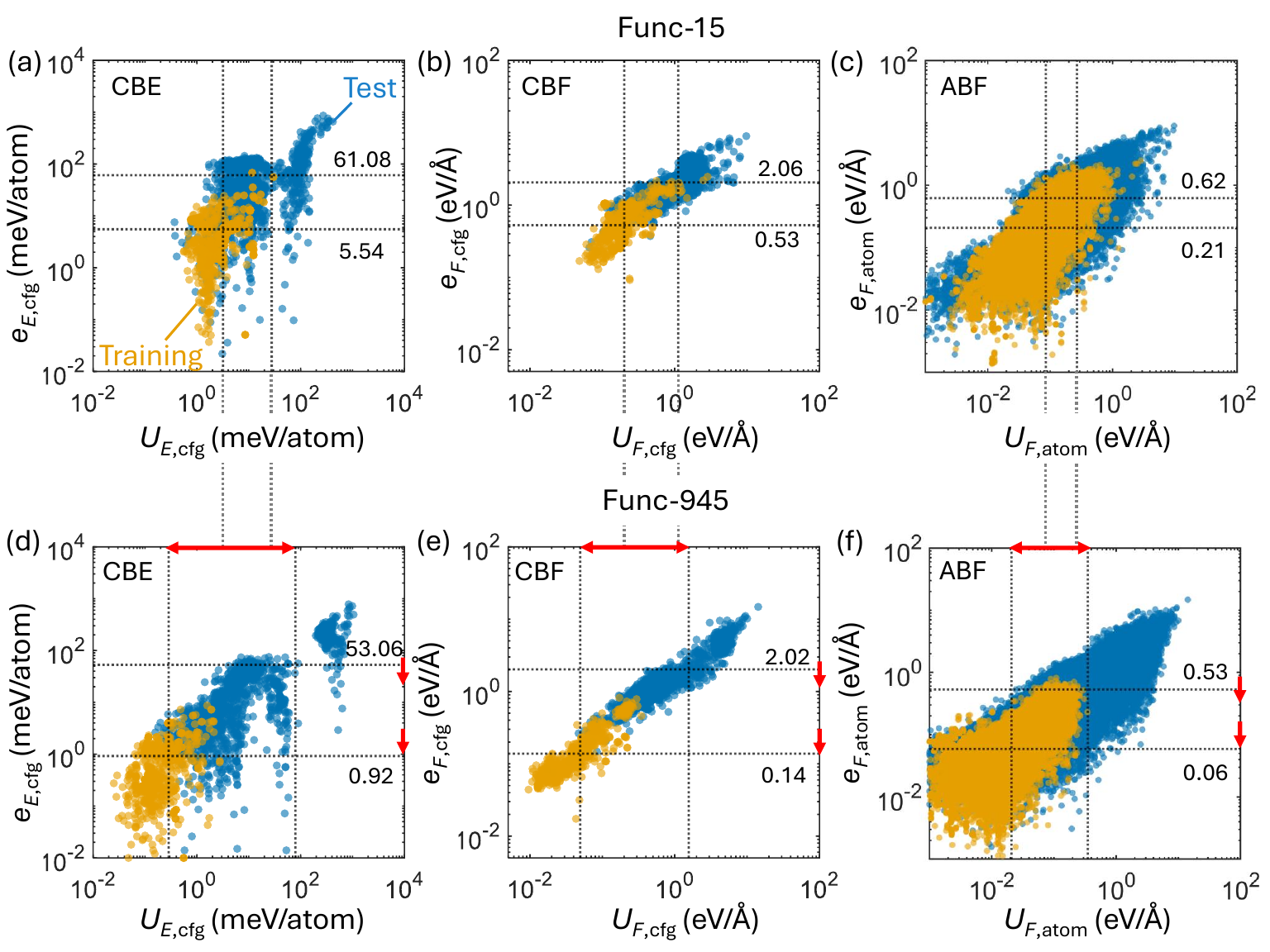}
\caption{\textbf{Correlation between model error and uncertainty evaluated using ensemble learning with ACE models.} 
\textbf{(a--c)} Func-15 models (15 functions) with $F_\mathrm{RMSE} = 171.62$\,meV/\si{\angstrom} and $E_\mathrm{RMSE} = 9.05$\,meV/atom. 
\textbf{(d--f)} Func-945 models (945 functions) with $F_\mathrm{RMSE} = 51.37$\,meV/\si{\angstrom} and $E_\mathrm{RMSE} = 1.46$\,meV/atom. 
Columns represent different uncertainty criteria: 
\textbf{(a, d)} configuration-based energy (CBE), 
\textbf{(b, e)} configuration-based force (CBF), and 
\textbf{(c, f)} atom-based force (ABF). 
Dashed lines show mean error and uncertainty values. 
Arrows indicate systematic performance shifts between Func-15 and Func-945 ensembles.}
    \label{fig2-ensemble-correlation}
\end{figure}

To illustrate the impact of model accuracy on UQ, we first present two ACE models at opposite ends of the basis-set complexity: the compact Func‑15, which uses just 15 basis functions, and the expansive Func‑945, which employs 945 basis functions. We evaluate three UQ metrics: CBE (Fig.\ref{fig2-ensemble-correlation}a,d), CBF (Fig.\ref{fig2-ensemble-correlation}b,e), and ABF (Fig.\ref{fig2-ensemble-correlation}c,f). The results in Fig.~\ref{fig2-ensemble-correlation} reveal three key observations. First, CBE demonstrates weak error correlations for both models, with Func-945 showing only slight improvement. Second, both force-based metrics (CBF and ABF) achieve substantially stronger correlations, where CBF’s superior performance stems from its integration of structural information across all atoms in a configuration. Third, while increased model complexity significantly reduces training-set errors and uncertainties, test-set performance remains relatively unaffected, as indicated by the dashed lines and arrows in Fig.\ref{fig2-ensemble-correlation}. This persistent gap reflects the test data’s OOD nature and the growing separation between training and test distributions as models become more accurate.

\begin{figure}[!ht]
    \centering
    \includegraphics[width=1\linewidth]{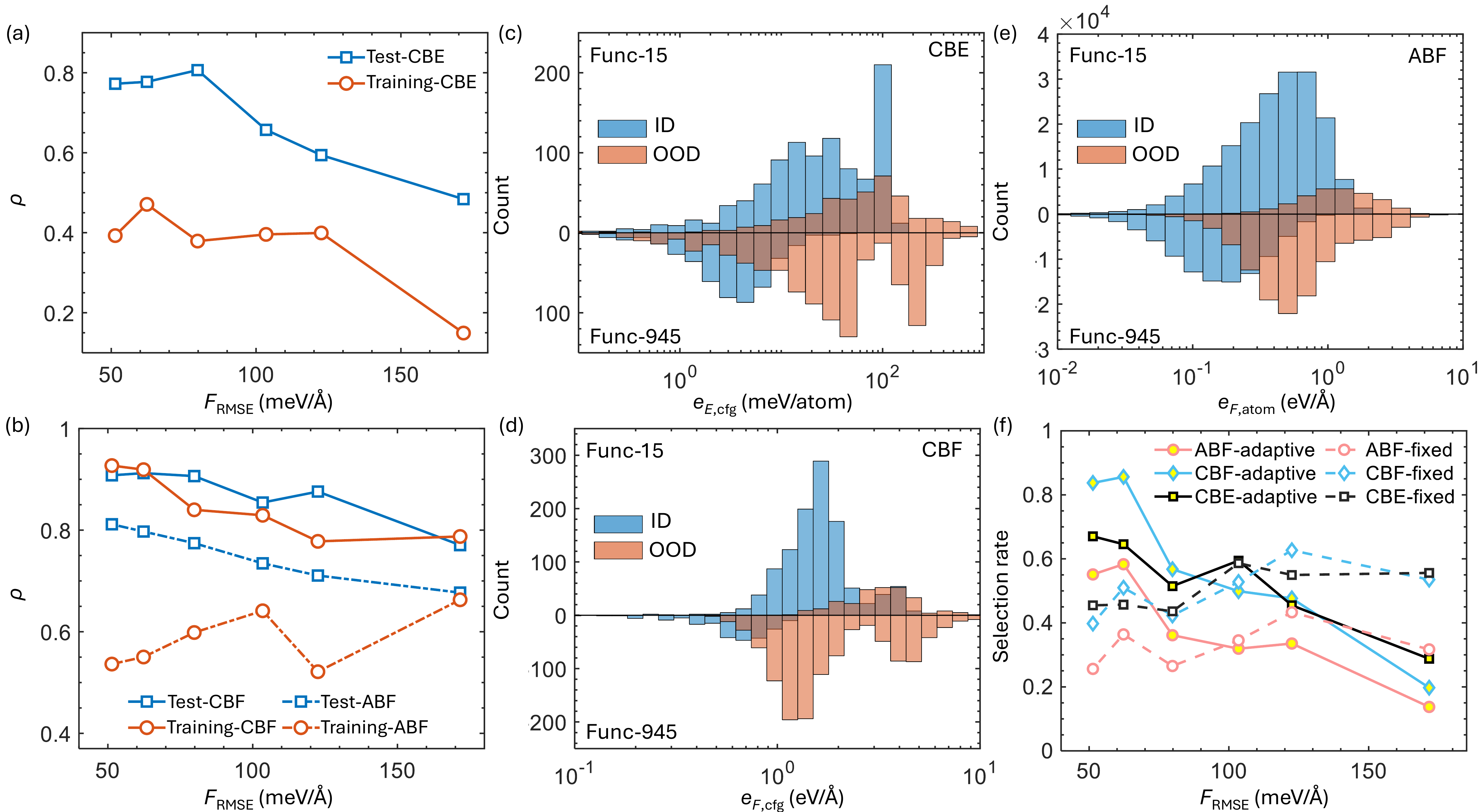}
    \caption{\textbf{Uncertainty quantification performance in ensemble learning: role of model accuracy.}
    \textbf{(a, b)} Spearman’s $\rho$ (uncertainty-error correlation) vs. force RMSE ($F_\mathrm{RMSE}$) for training/test data across three criteria: configuration-based energy (CBE), force (CBF), and atomic force (ABF).
    \textbf{(c–e)} Comparison between Func-15 and Func-945 models for in-distribution (ID) and out-of-distribution (OOD) detection across CBE, CBF, and ABF criteria for the test data.
    \textbf{(f)} Selection rate vs. $F_\mathrm{RMSE}$ using adaptive/fixed thresholds for all criteria.}
    \label{fig3-ensemble-spearman-detection}
\end{figure}

To systematically evaluate the impact of model accuracy, we compute Spearman’s rank correlation coefficient ($\rho$), a nonparametric measure of how closely the ordering of predicted uncertainties matches the ordering of observed errors, across models with progressively lower force root‑mean‑square error ($F_\mathrm{RMSE}$). Fig.\ref{fig3-ensemble-spearman-detection}a demonstrates that for the CBE criterion, correlation strength increases monotonically with increase in model accuracy for both training and test datasets, showing particularly dramatic increase in test data. The CBF criterion (Fig.\ref{fig3-ensemble-spearman-detection}b, solid line) shows analogous accuracy dependence while achieving substantially stronger correlations than CBE. Notably, the ABF criterion (dashed line) reveals divergent behavior: test data correlations increase steadily with accuracy, training set correlations remain consistently low ($\rho < $ 0.7) and show no systematic relationship with model accuracy.
Three fundamental insights emerge from this analysis. First, force-based criteria (CBF and ABF) universally surpass the energy-based CBE in correlation strength. Second, CBF consistently outperforms ABF. Third, and most significantly, test data correlations not only benefit more from improved model accuracy than training data, but also maintain superior absolute correlation strength across all accuracy levels. These findings collectively establish that robust UQ requires both careful metric selection and ongoing model refinement, with force-based configuration-level analysis delivering optimal performance for practical applications involving defection of novel configurations or LAEs.

The primary goal of UQ is to detect unseen configurations and LAEs. We derive UQ thresholds for CBE ($\varepsilon_{E,\mathrm{cfg}}$), CBF ($\varepsilon_{F,\mathrm{cfg}}$), and ABF ($\varepsilon_{F,\mathrm{atom}}$) (see~\nameref{methods}) to flag OOD configurations and LAEs. Applying these thresholds to the combined C, E, and F test sets (Fig.~\ref{fig1-pca}), we identify OOD configurations using CBE (Fig.~\ref{fig3-ensemble-spearman-detection}c) and CBF (Fig.~\ref{fig3-ensemble-spearman-detection}d), and detect OOD LAEs using ABF (Fig.~\ref{fig3-ensemble-spearman-detection}e) for both the Func-15 and Func-945 models. The Func-15 model selects very few new configurations or LAEs, classifying most test cases as ID despite high errors. In contrast, the more accurate Func-945 model flags a substantial fraction of new configurations and LAEs, due to the clearer separation between training and test data (Fig.~\ref{fig2-ensemble-correlation}). Fig.~\ref{fig3-ensemble-spearman-detection}f illustrates how the selection rate of each criterion scales with model accuracy, defined as the fraction of flagged configurations (relative to total test configurations) or LAEs (relative to total test-set atoms). Higher model accuracy consistently yields more flagged items. Notably, at comparable accuracy levels, CBF outperforms CBE in detecting novel configurations, a trend particularly evident for the highest-fidelity ACE models.

A key remaining question concerns the relative performance of adaptive versus fixed thresholds for OOD detection. We assess this by applying the mean thresholds of our three criteria (CBE, CBF, and ABF) across different $F_\mathrm{RMSE}$ levels (Fig.~\ref{fig:S1}) as fixed thresholds to evaluate selection rates. As shown by the dashed lines in Fig.~\ref{fig3-ensemble-spearman-detection}f, fixed thresholds exhibit selection rates with minimal dependence on $F_\mathrm{RMSE}$. While both approaches demonstrate similar selection rates at $F_\mathrm{RMSE} = 100\,\mathrm{meV/\AA}$, fixed thresholds identify more configurations/LAEs below this value and fewer above it. However, while fixed thresholds may select more configurations/LAEs at low $F_\mathrm{RMSE}$, this does not necessarily indicate better OOD detection accuracy. These findings collectively demonstrate the superior reliability of adaptive thresholds for OOD detection. 

We also evaluate how ensemble size affects the detection of novel configurations and LAEs. Using our most-accurate ACE model (Func-945) with ensemble sizes ranging from 3 to 30 models, Fig.~\ref{fig:S2}a shows that force-based metrics (CBF and ABF) exhibit strong ensemble-size dependence, while CBE remains relatively stable. All three criteria achieve consistent selection rates only when the ensemble contains $\geq$10 models, which is twice the conventional five-model standard \cite{Erhard2024}.
To understand this dependence, we compute the Spearman correlation $\rho$ between prediction error and uncertainty for both training (A+B) and test (C+E+F) sets (Fig.~\ref{fig:S2}b,c). The fluctuating $\rho$ values reveal no systematic trend with ensemble size, indicating Spearman's $\rho$ alone cannot explain the detection trends. Analysis of prediction errors (Fig.~\ref{fig:S2}d--i) shows larger ensembles simultaneously increase test-set errors while decreasing training-set errors. This growing train-test divergence enhances novel configuration/LAE detection, an effect distinct from model accuracy effects in Fig.~\ref{fig2-ensemble-correlation}.
Moreover, larger ensembles provide two key advantages: (1) increased mean test-set uncertainty (Fig.~\ref{fig:S2}j), and (2) reduced novelty-detection thresholds $\varepsilon$ (Fig.~\ref{fig:S2}k--m), except for CBE (Fig.~\ref{fig:S2}m). These lower thresholds enable more OOD flagging, fully explaining the rising selection rates in Fig.~\ref{fig:S2}a.

\subsection*{D-optimality criterion and MaxVol algorithm}

In our analysis of the D‑optimality criterion, we use the extrapolation grade \(\gamma\) computed via the MaxVol algorithm for UQ (see \nameref{methods}). Analogous to the ensemble approach, we derive \(\gamma_{\mathrm{cfg}}\) and \(\gamma_{\mathrm{atom}}\) to assess the uncertainty of entire configurations and individual atoms, respectively.

\begin{figure}[!ht]
    \centering
    \includegraphics[width=0.99\linewidth]{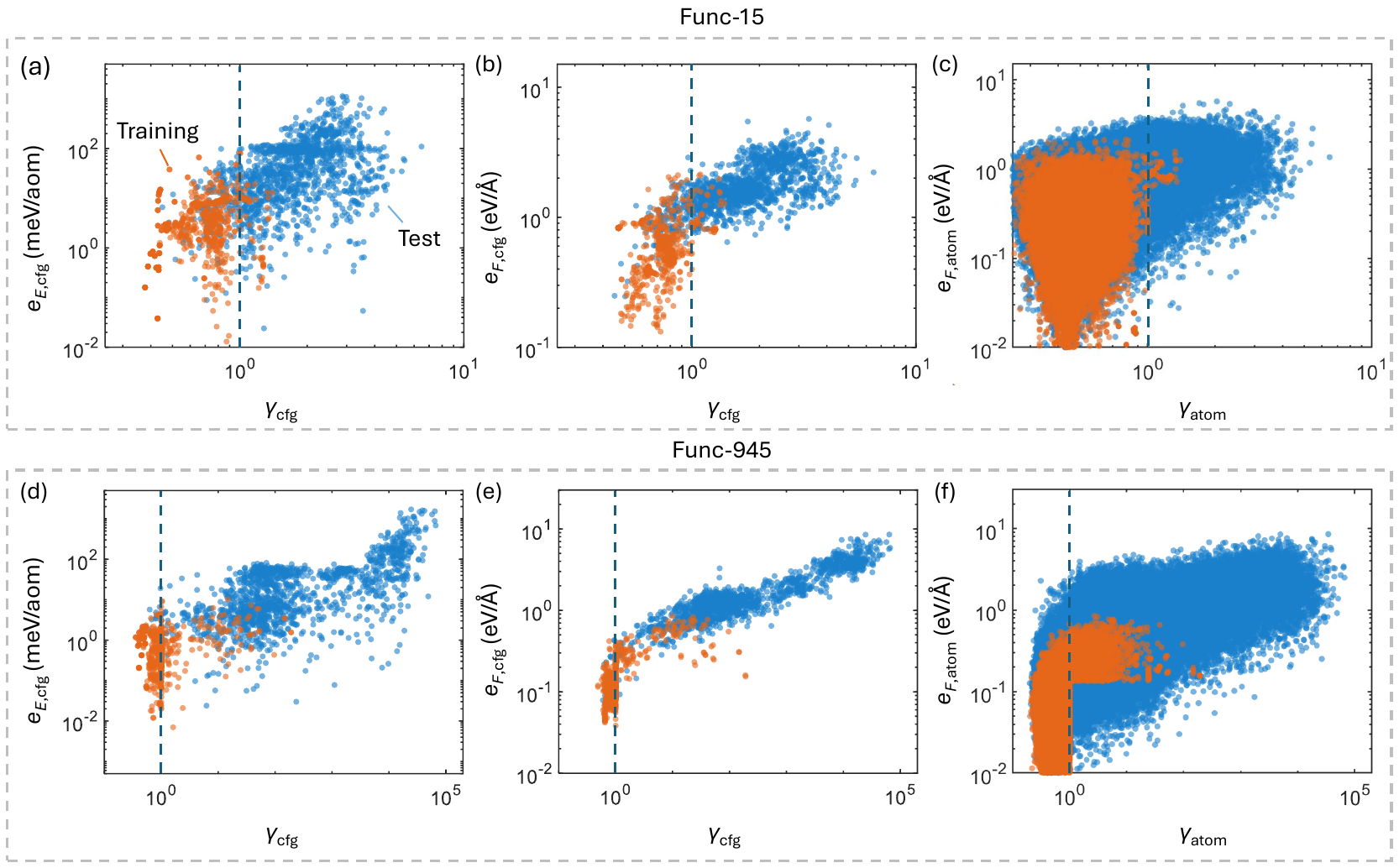}
    \caption{\textbf{Prediction Errors vs. Extrapolation Grade ($\gamma$) in ACE Models.}
    Results are shown for two models: \textbf{(a--c)} Func-15 and \textbf{(d--f)} Func-945. \textbf{(a, d)} Configurational energy errors. \textbf{(b, e)} Configurational force errors. \textbf{(c, f)} Atomic-level force errors. Vertical dashed lines mark the extrapolation threshold ($\gamma = 1$).}
    \label{fig4-gamma-correlation}
\end{figure}

We first consider A + B as the training set with C + E + F for testing. Fig.\ref{fig4-gamma-correlation} presents extrapolation grades at both configuration and atom level ($\gamma_\mathrm{cfg}$ and $\gamma_\mathrm{atom}$), plotted against energy and force errors. The threshold $\gamma = 1$ (dashed line in the figure) separates ID ($\gamma < 1$) from OOD ($\gamma > 1$) regimes across both models. Our D-optimality analysis reveals distinct patterns in UQ when comparing the Func-15 and Func-945 models. The more accurate Func-945 model (panels d-f) shows significantly stronger error-grade correlations than Func-15 (panels a-c), consistent with ensemble method results in Fig.\ref{fig2-ensemble-correlation}. The range of $\gamma$ values also differs by orders of magnitude: Func-15 yields grades around $10^2$, whereas Func-945 reaches values near $10^6$, highlighting how higher model accuracy improves discrimination among configurations and LAEs. For both models, configurational energy errors (Fig.\ref{fig4-gamma-correlation}a,d) and force errors (Fig.\ref{fig4-gamma-correlation}b,e) remain random below $\gamma_\mathrm{cfg} = 1$ but increase markedly once $\gamma_\mathrm{cfg}$ exceeds 1. Overall, these results confirm that D‑optimality effectively identifies OOD configurations and that $\gamma_\mathrm{cfg}$ correlates more strongly with configuration force errors than with energy errors, consistent with the ensemble learning trends shown in Fig.\ref{fig3-ensemble-spearman-detection}. At the atomic level (Fig.\ref{fig4-gamma-correlation}c,f), $\gamma_\mathrm{atom}$ identifies more OOD LAEs in the Func-945 case, yet the per-atom force errors show only a weak dependence on $\gamma_\mathrm{atom}$. Notably, many atoms with $\gamma_\mathrm{atom} > 1$ exhibit very low errors, indicating potential extrapolation capability of the MLIP. These results collectively establish D-optimality as a robust method for configuration-level UQ, while revealing inherent limitations in atomic-level analysis.

\begin{figure}[!ht]
    \centering
    \includegraphics[width=1\linewidth]{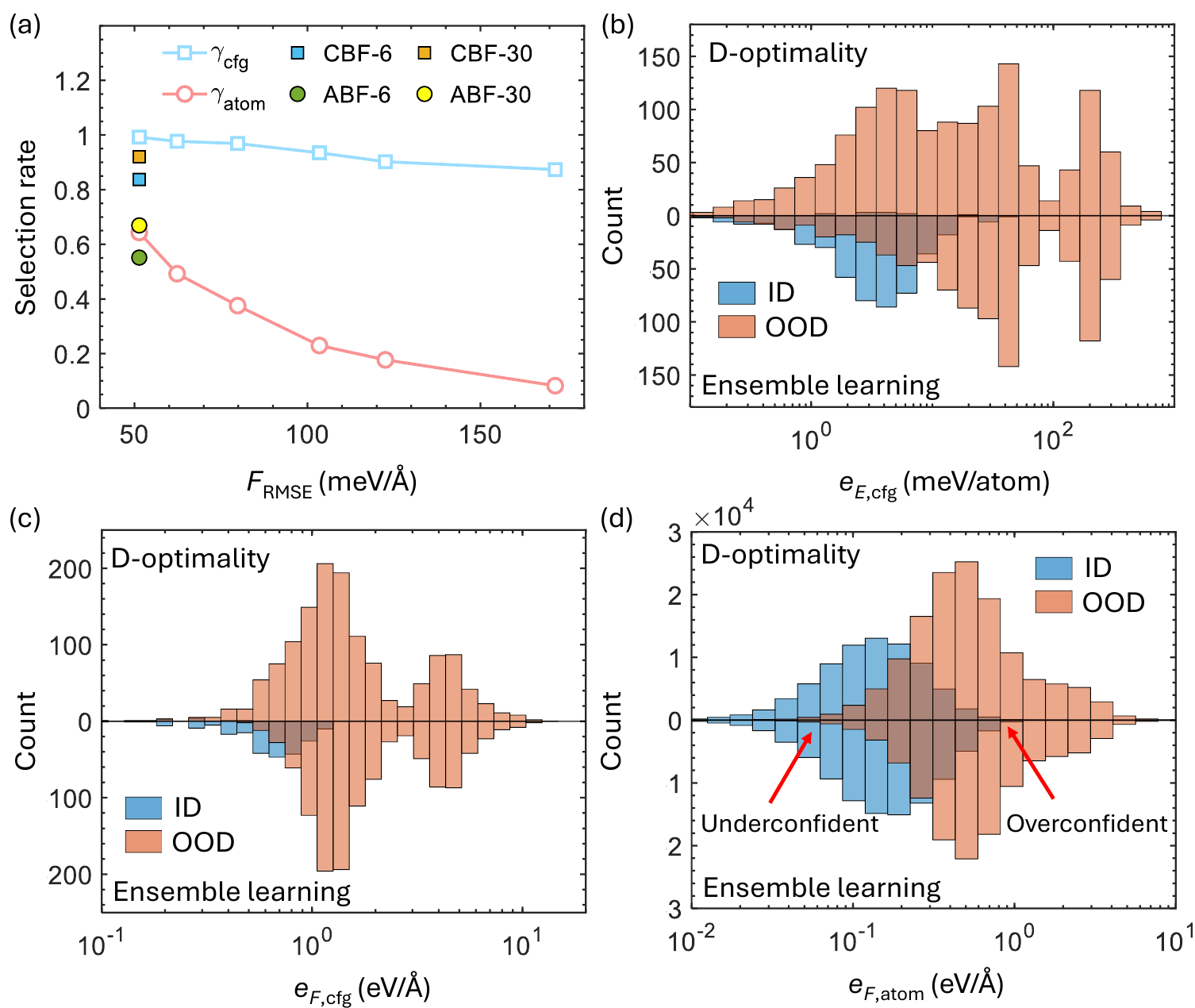}
    \caption{\textbf{Comparison of ensemble learning and D-optimality methods.}
    \textbf{(a)} OOD detection rate versus $F_\mathrm{RMSE}$.
    \textbf{(b--d)} Classification performance for in-distribution (ID) and out-of-distribution (OOD) configurations/atoms.
    In panel \textbf{(d)}, left arrow indicate underconfident predictions (low-error LAEs misidentified as OOD), right arrow show overconfident cases (high-error LAEs misclassified as ID).}
    \label{fig5-gamma-detection}
\end{figure}

We then compare OOD detection performance between ensemble learning and D-optimality approaches in Fig.~\ref{fig5-gamma-detection}. The solid lines in Fig.~\ref{fig5-gamma-detection}a demonstrate that D-optimality achieves consistently high configuration-level detection ($>90\%$) across all model accuracies, while LAE detection improves from $\sim$5\% to $\sim$70\% with increasing accuracy. Compared to both 6-member and 30-member ensemble results, D-optimality shows superior configuration-level detection and comparable atomic-level performance, despite requiring only a single ACE model. This reveals D-optimality's dual advantages of more conservative detection and greater computational efficiency relative to ensemble methods.

The detailed comparison between ensemble learning and D-optimality is shown in Fig.~\ref{fig5-gamma-detection}b--d, contrasting their ability to identify ID and OOD configurations/LAEs in the combined C+E+F test set using Func-945 potentials. D-optimality demonstrates superior detection performance, flagging over $99\%$ of test configurations as OOD (upper panels in Fig.~\ref{fig5-gamma-detection}b,c). In contrast, ensemble methods miss significant fractions of high-error cases: the energy-based ensemble overlooks $\sim$$33\%$ and the force-based ensemble $\sim$$16\%$, incorrectly labeling them as ID (lower panels).
At the atomic level (Fig.~\ref{fig5-gamma-detection}d), D-optimality identifies $64\%$ of atoms as OOD LAEs versus $55\%$ for ensembles, demonstrating more comprehensive local environment sampling. HHowever, both approaches exhibit characteristic limitations: they incorrectly classify high-error atomic sites (up to \qty{1}{eV/\angstrom}) as ID (demonstrating overconfidence) while flagging low-error sites (\qty{0.05}{eV/\angstrom}) as OOD (showing underconfidence), as highlighted by the arrows. This reflects the fundamental challenge of atomic-level active learning compared to whole-configuration sampling. Neither method achieves perfect discrimination - both systematically miss critical high-error sites while oversampling well-predicted regions, leading to inefficient computational resource allocation that undermines overall sampling efficiency.

\subsection*{Influence of data heterogeneity}\label{data-hetero}

To probe the limitations of ensemble learning and D optimality on structurally heterogeneous data, we devise a stringent scenario. The training set consists of 30 elastic deformation configurations (dataset A) and 30 nanospheres (dataset D), while datasets B, C, E, and F serve as the test set. This arrangement echoes the neighborhood mode of MLIP 3’s active learning framework\cite{Podryabinkin2023}, in which vacuum‑embedded clusters are constructed so that novel LAEs occupy the cluster center. By applying both UQ methods in this context, we uncover their respective blind spots and derive practical lessons for optimizing active‑learning protocols to heterogeneous training sets.

\begin{figure}[!ht]
    \centering
    \includegraphics[width=1\linewidth]{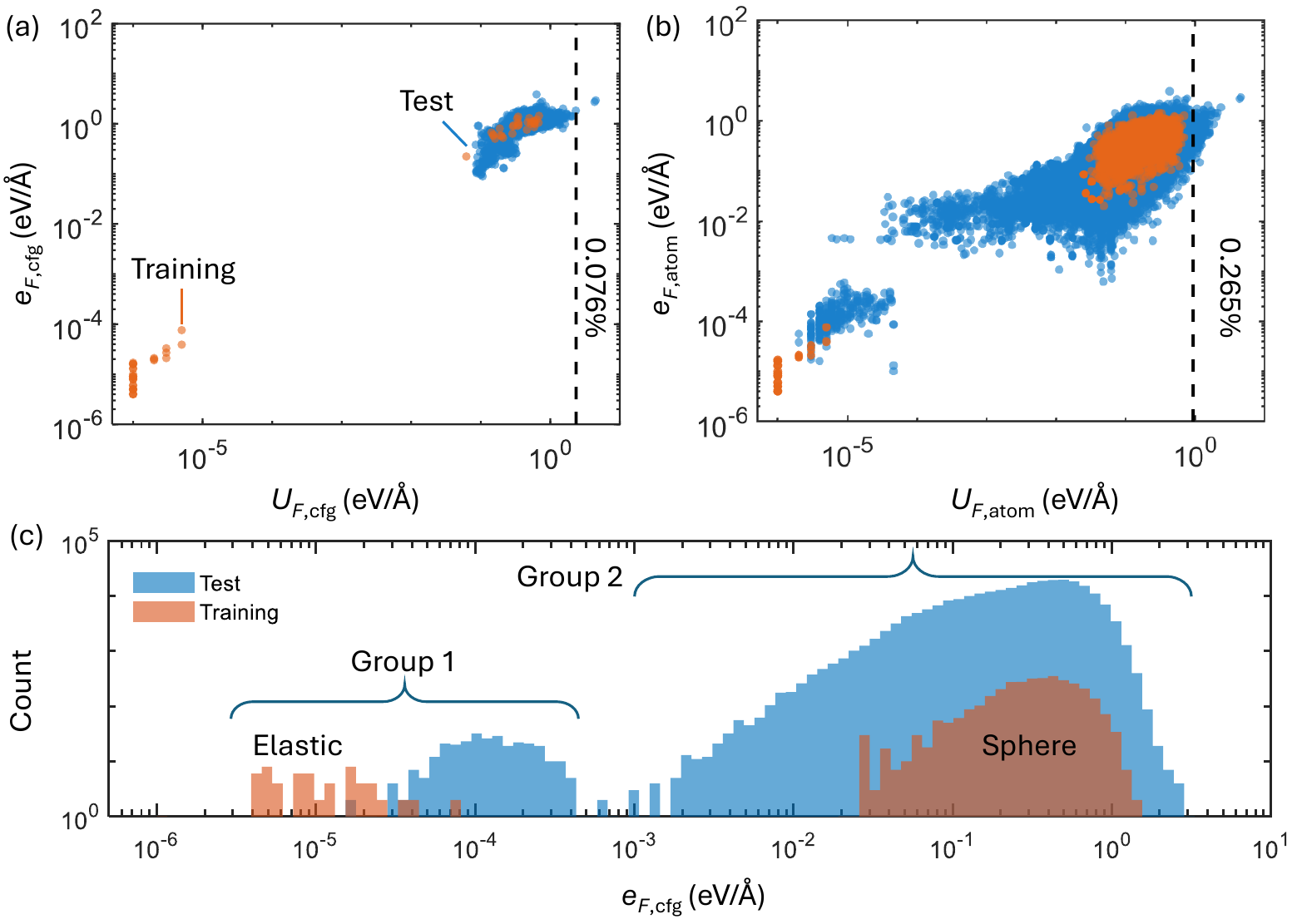}
    \caption{\textbf{Limitations of ensemble learning for datasets containing subsets with heterogeneous complexity.} \textbf{(a)} Force error vs. uncertainty at the configurational level. The vertical dashed line marks the critical uncertainty threshold for identifying OOD configurations. Only $0.076\%$ of test configurations are flagged as OOD. \textbf{(b)} Force error vs. uncertainty at the atomic level, where just $0.265\%$ of test atoms are classified as OOD. \textbf{(c)} Distribution of atomic force errors in training and test datasets, revealing two distinct subgroups with differing complexity in both datasets (elastic and sphere for the training set, and group 1 and 2 for the test set). }
   \label{fig6-limitation-ensemble}
\end{figure}

All ensemble learning uncertainty calculations employ Func-945 models. Fig.~\ref{fig6-limitation-ensemble} reveals a fundamental paradox in ensemble-based UQ: despite strong force error-uncertainty correlations at both configurational (Fig.~\ref{fig6-limitation-ensemble}a) and atomic (Fig.~\ref{fig6-limitation-ensemble}b) levels, the method fails catastrophically for novelty detection. The detected OOD fractions (only $0.076\%$ of configurations and $0.265\%$ of atoms, corresponding to data points beyond the dashed uncertainty thresholds) represent complete failure, since the entire test set should be identified as OOD by design. This conclusion is unequivocal given that the training set contained just two structural motifs (elastically deformed bulk structures and BCC nanospheres), while the test set consists entirely of different defect-bearing configurations.

This critical failure originates from the training data's intrinsic heterogeneity. Fig.~\ref{fig6-limitation-ensemble}c reveals that the training-set force errors exhibit bimodal distribution: one mode corresponds to easily predicted elastic-deformation configurations, while the other reflects the inherently more complex nanosphere surface environments. A single global uncertainty threshold, forced to accommodate both regimes, becomes dominated by the high-error nanosphere population and consequently sets an excessively high threshold for the elastic-deformation cases. 
The test set replicates this bimodal structure, with clusters centered near $10^{-4}$\,eV/\si{\angstrom} (Group 1) and $10^{-1}$\,eV/\si{\angstrom} (Group 2). As a result, the unified cutoff even fails to identify high-error Group 2 sites as OOD. This prevalence of false negatives in the high-error regime not only compromises UQ's reliability for active learning and adaptive sampling but also exposes the fundamental limitation of single-threshold methods when applied to multimodal error distributions.

\begin{figure}[!ht]
    \centering
    \includegraphics[width=1\linewidth]{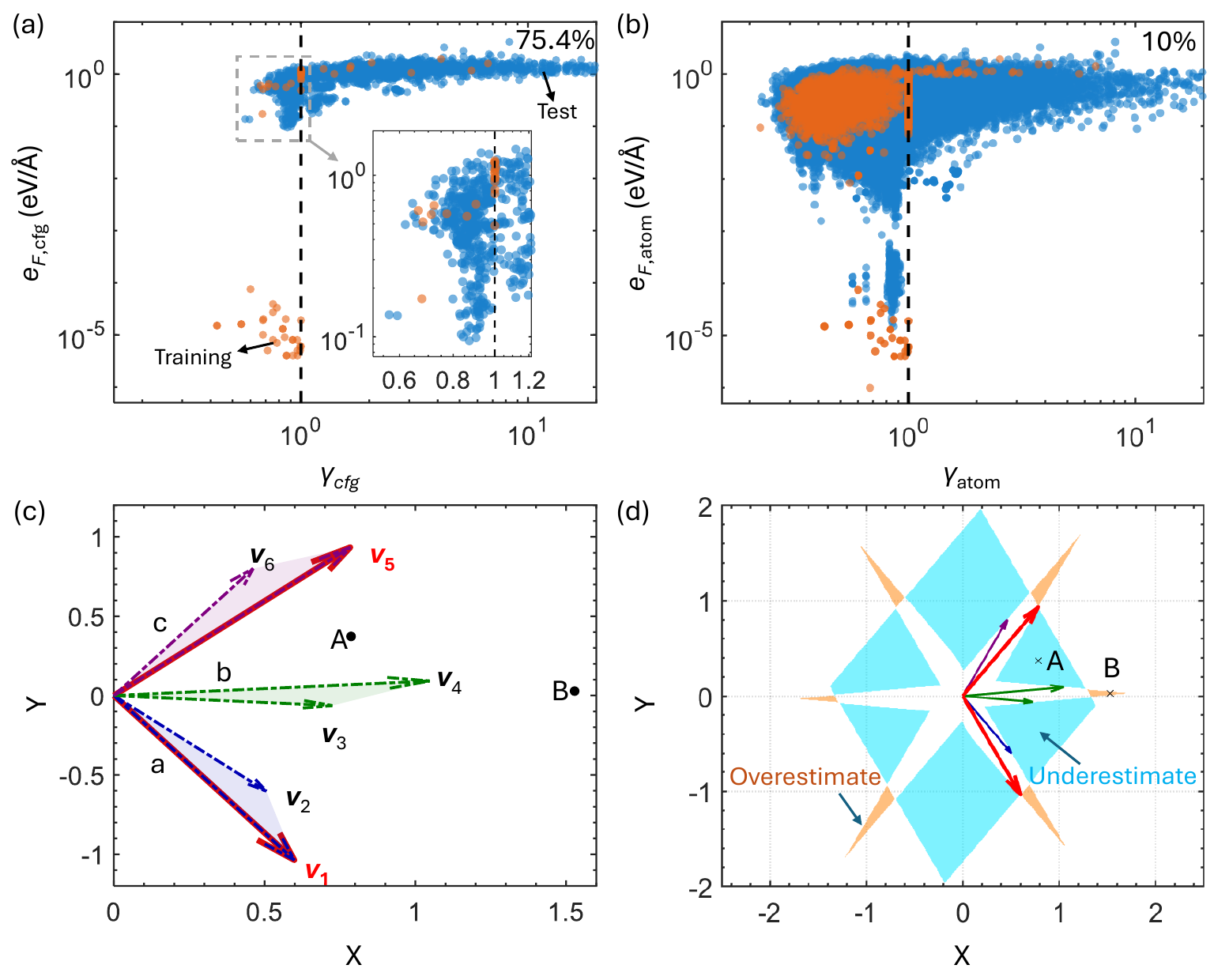}     \caption{\textbf{Limitations of D-optimality for datasets with subsets of heterogeneous complexity.} \textbf{(a)} Force error vs. extrapolation grade ($\gamma_\mathrm{cfg}$) at the configurational level. The vertical dashed line marks the critical grade ($\gamma_\mathrm{cfg} > 1$) for flagging OOD configurations. The inset zooms in on the data within the dashed rectangle. $75.4\%$ of test configurations are labeled as OOD. \textbf{(b)} Force error vs. extrapolation grade ($\gamma_\mathrm{atom}$) at the atomic level, with only $10\%$ of test atoms identified as OOD. \textbf{(c)} Three groups of vectors, each containing two vectors as an active set. Vectors $\mathbf{v_1}$ and $\mathbf{v_5}$ form the active set for the combined data. Test points A and B are highlighted. \textbf{(d)} Overestimated (false OOD) and underestimated (missed OOD) regions when calculating the extrapolation grade ($\gamma$) using the combined dataset.}
   \label{fig7-limitation-gamma}
\end{figure}

Using Func‑945 models, we compute D‑optimality extrapolation grades by training on 30 configurations each from datasets A and D and testing on the combined B + C + E + F set. Figure\ref{fig7-limitation-gamma} compares force errors against these grades at both the configurational and atomic scales. At the configuration level in Fig.\ref{fig7-limitation-gamma}a, D‑optimality flags $75.4\%$ of test structures as OOD, improving on the ensemble method (Fig.\ref{fig6-limitation-ensemble}) yet still inadequate given that every test configuration is, by design, OOD. At the atomic level in Fig.\ref{fig7-limitation-gamma}b, only $10\%$ of local environments are detected as OOD. Compared with the ensemble results in Fig.\ref{fig6-limitation-ensemble}, extrapolation grades show much better selection rates but weaker correlation with force errors. Moreover, the $\gamma$ values span just 0.1 to 10, a dramatically narrower range than the $10^6$ observed for the homogeneous A + B training set as shown in Fig.\ref{fig4-gamma-correlation}. These observations demonstrate that structural heterogeneity constrains both the magnitude and the predictive reliability of D‑optimality grades.

To further elucidate the limitations of D-optimality and MaxVol algorithm, we present a simplified two-dimensional example in Fig.\ref{fig7-limitation-gamma}c demonstrating the MaxVol algorithm’s active set selection and extrapolation grade calculation, where three distinct non-overlapping subsets (a, b, c) with respective active sets ($\mathbf{v_1}$,$\mathbf{v_2}$), ($\mathbf{v_3}$,$\mathbf{v_4}$), and ($\mathbf{v_5}$,$\mathbf{v_6}$) combine to form a new active set ($\mathbf{v_1}$,$\mathbf{v_5}$). This analysis reveals critical inconsistencies in extrapolation grade determination: while point A appears ID ($\gamma_{15} = 0.76$) and point B OOD ($\gamma_{15} = 1.17$) in the combined dataset, examination of individual subsets shows the opposite behavior: point A consistently demonstrates OOD character ($\gamma_{12} = 6.36$, $\gamma_{34} = 2.41$, $\gamma_{56} = 2.34$) while point B is clearly ID ($\gamma_{34} = 0.88$) as it belongs to subset b. A comprehensive regional scan (Fig.\ref{fig7-limitation-gamma}d) further demonstrates that grade calculations based on the combined dataset overwhelmingly tend toward underestimation, with only rare cases of overestimation, as exemplified by points A and B respectively. These results highlight a core weakness of MaxVol: it targets only the extreme vertices of training dataset and ignores interior points. Novel data that lie within this hull receive low $\gamma$ values, remain unselected, and leave large regions of configuration space unsampled, ultimately constraining the reach of D-optimality based active learning in MLIP development.

\subsection*{Improved D-optimality approach}

\begin{figure}[!ht]
    \centering
    \includegraphics[width=0.9\linewidth]{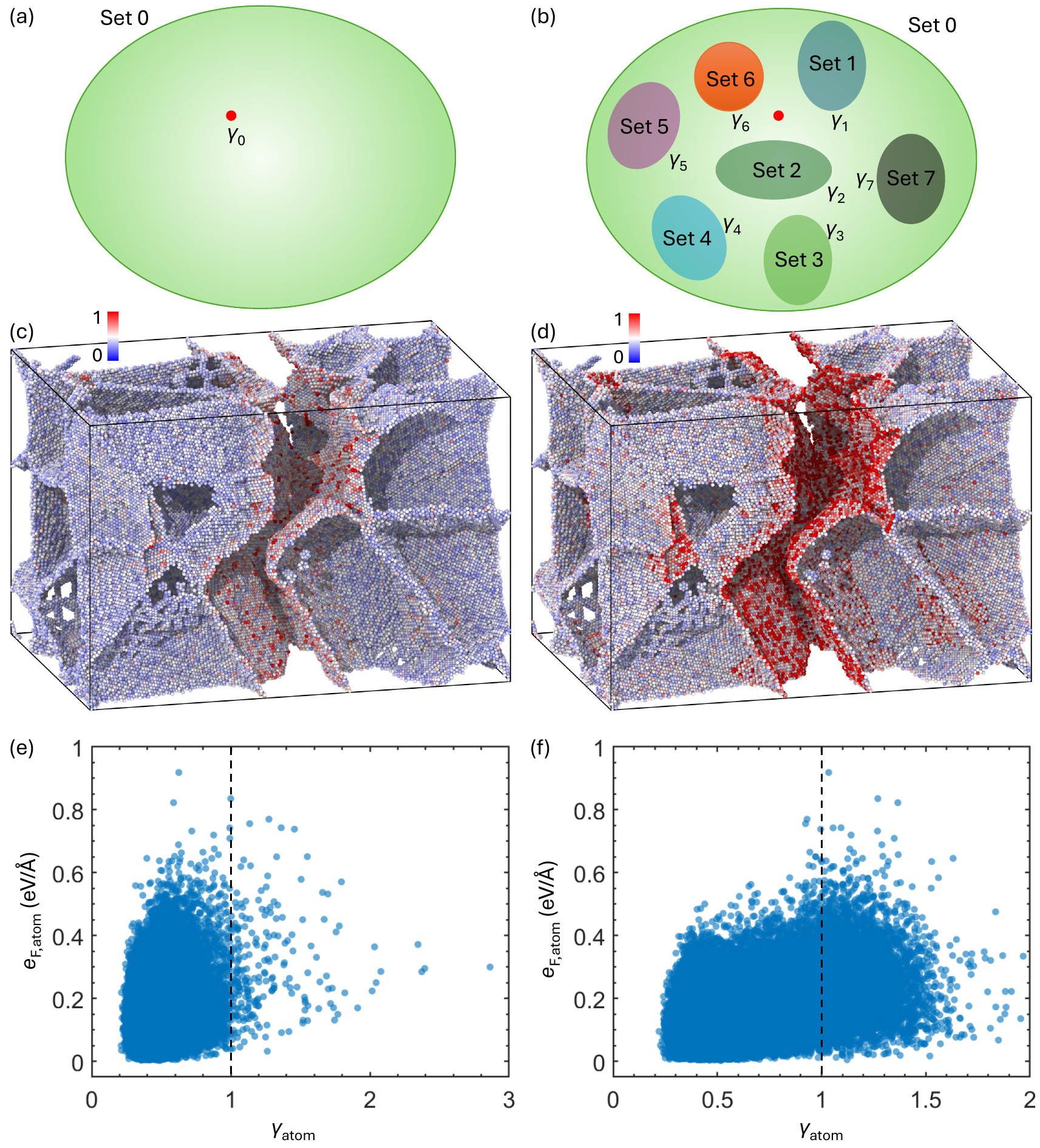}
    \caption{\textbf{A new D-optimality approach for uncertainty quantification.}
    Schematic illustrations compare (\textbf{a}) the original D-optimality with (\textbf{b}) our clustering-enhanced local D-optimality method. Atomistic configurations of a fractured tungsten (W) polycrystal are shown, with atomic colors indicating the extrapolation grade ($\gamma_\mathrm{atom}$) computed using (\textbf{c}) the original D-optimality and (\textbf{d}) its improved variant. Scatter plots demonstrate the correlation between atomic force errors and extrapolation grades for (\textbf{e}) the original and (\textbf{f}) the refined approach.}
   \label{fig8-application}
\end{figure}

To overcome the D-optimality limitations revealed in Fig.\ref{fig7-limitation-gamma}, we propose a clustering-enhanced local D-optimality approach that significantly improves uncertainty quantification for structurally diverse datasets, as shown in Fig.\ref{fig8-application}. The key insight stems from recognizing that conventional single-grade calculations ($\gamma_\mathrm{cfg}$ or $\gamma_\mathrm{atom}$) systematically underestimate novelty in heterogeneous dataset (Fig.\ref{fig7-limitation-gamma}), prompting our modified algorithm to instead compute subset-specific grades ($\gamma_{\mathrm{cfg},i}$ or $\gamma_{\mathrm{atom},i}$) and select their minimum as the final metric, a strategy that simultaneously prevents both underestimation by combined datasets and overestimation from individual subsets ( as shown in Fig.~\ref{fig8-application}a,b). This approach proves particularly useful for identifying transitional configurations between distinct structural regimes, as demonstrated by the point A in Fig.\ref{fig7-limitation-gamma}c,d: where traditional methods would erroneously classify this boundary-spanning environment as ID, our minimum-grade criterion correctly flags it as OOD, thereby capturing crucial yet easily overlooked atomic environments that are essential for developing truly comprehensive MLIP.

To validate our clustering‑enhanced D‑optimality approach, we apply it to the W dataset sourced from Ref.\cite{byggmastar2020gaussian}. This dataset comprises a diverse set of pre‑labeled subgroups, including distorted BCC unit cells, FCC and HCP crystals, high‑temperature BCC phases, vacancies, self‑interstitials, surface configurations, liquids, and others. Rather than using the original DFT energies and forces, we employ predictions from the universal NEP89 potential\cite{liang2025nep89} to label all structures, thereby enabling the calculation of true errors for large scale configurations. For each pre‑labeled subgroup, we train a dedicated ACE model and assemble its active set. We then compute the extrapolation grade $\gamma$ for every atom with respect to each active set and assign each atom the minimum $\gamma$ value across all subgroup models as its final extrapolation grade. We test this procedure on a fractured polycrystal model from our recent work\cite{shuang2025modeling}. As shown in Fig.\ref{fig8-application}, the original D‑optimality method (Fig.\ref{fig8-application}c) flags only a few fracture‑surface atoms as OOD, despite leaving many high‑error ID atoms undetected (Fig.\ref{fig8-application}e). By contrast, our clustering‑enhanced version (Fig.\ref{fig8-application}d) correctly identifies a much larger set of fracture‑surface atoms as OOD, all with $\gamma > 1$. Crucially, Fig.\ref{fig8-application}f confirms that these newly detected atoms consistently exhibit higher force errors, demonstrating the superior reliability of our method for UQ.

\section*{DISCUSSION}

Our study reveals consistent principles and key distinctions between both UQ methods. For ensemble learning, we establish three critical findings. First, force-based criteria (CBF/ABF) show superior error-uncertainty correlations compared to energy-based metrics (CBE), with configuration-level analysis proving more reliable than atomic-level assessment. Second, model accuracy plays a crucial role in effective novelty detection. Third, robust detection requires larger ensembles of at least 10 models for stable performance.
These principles also apply to D-optimality approaches, where configuration-level metrics similarly outperform atomic-level analysis in error correlation. However, a key difference emerges regarding accuracy dependence: atomic-level D-optimality detection shows strong sensitivity to model accuracy, while configuration-level performance remains largely accuracy-independent.
Both methods exhibit qualitatively similar novelty identification behavior, with D-optimality offering a more conservative and computationally efficient alternative to ensemble learning. While increasing MLIP count can improve ensemble detection capability, this comes at substantial computational cost during both training and inference.
We therefore recommend D-optimality as the preferred acquisition criterion. When unavailable (e.g., for universal MLIPs), ensemble methods must incorporate force-based analyses, high-fidelity models, and sufficiently large ensemble sizes (minimum 10 models) to ensure adequate performance.

Critically, our analysis reveals fundamental limitations in both ensemble and D-optimality UQ methods when handling heterogeneous training data. These approaches systematically fail to properly quantify uncertainty across multimodal distributions, leading to unreliable novelty detection. This failure stems from their inability to simultaneously accommodate diverse atomic environments.
Yet this heterogeneity is unavoidable in practice. Proper MLIP training sets must encompass the complete spectrum of atomic environments found in real materials, including surfaces, interfaces, point defects, and bulk polymorphs across multiple space groups \cite{poul2025automated}. They must also incorporate extreme configurations like isolated atoms, dimers at varying separations, and collision geometries relevant to radiation-damage cascades \cite{byggmastar2020gaussian}. The RANDSPG algorithm's material-agnostic approach, enumerating all 230 space groups with random primitive cells of 3-10 atoms \cite{Poul2023}, further demonstrates this inherent diversity. For high-entropy alloys, the challenge compounds as structural and chemical diversity interact in ways not yet fully understood.
This unavoidable heterogeneity creates a fundamental tension: while current UQ methods work well for near-homogeneous data, they break down for the complex, multimodal distributions required for robust MLIP development. Our results expose this critical gap in the workflow of MLIP development, where inadequate UQ leads to persistent undersampling of precisely those atomic environments that are most informative yet most challenging to model.

Our findings have significant implications for on-the-fly active learning of LAEs in large-scale simulations, where atom-based UQ is required. In the standard MLIP-3 and \textit{pacemaker} workflows, a spherical cluster around each candidate “core” atom is extracted, enclosed in vacuum layers, and appended to the training set. However, this practice inadvertently incorporates surface atoms that are irrelevant to bulk-focused simulations. Because these extreme surface configurations substantially enlarge the envelope of active set in the MaxVol algorithm as illustrated in Fig.\ref{fig7-limitation-gamma}a,b, the extrapolation grade underestimates the novelty of true bulk environments in the following active learning; genuinely new local structures are misclassified as ID simply because they are less exotic than the spurious surface atoms. Consequently, the original extrapolation-grade criterion renders on-the-fly active learning in MLIP-3 and \textit{pacemaker} ineffective for generating truly local, bulk-specific MLIPs. A simple remedy is to construct the active set using only the core atoms, thereby excluding those with artificially truncated coordination. Alternatively, one can fill the vacuum region via empirical interatomic potential-guided grand-canonical Monte Carlo (EIP-GCMC) and retain only the lowest-energy configurations\cite{shuang2025modeling}. Both strategies preserve structural relevance to the target simulation, prevent dilution of the uncertainty metric by spurious surfaces, and restore the extrapolation grade’s sensitivity to genuinely novel local structures.

Yet the most reliable ensembler learning and D-optimality based UQ must be performed locally, gradually and independently for each candidate environment during on-the-fly active learning. Hodapp et al. recently exemplified this approach by embedding an isolated screw dislocation in BCC metals or partial dislocation in FCC metals into a fully periodic supercell while excluding all atomic environments outside the defect core\cite{Hodapp2020, Mismetti2024}. By calculating the extrapolation grade solely within this narrowly defined region, their acquisition algorithm accurately identifies truly novel dislocation configurations and discards spurious outliers. The resulting MTP achieves remarkably low fitting errors and accurately reproduces the Peierls barrier, demonstrating that a defect-centered, locality-preserving sampling strategy is essential for reliable active learning. If the initial training set is heterogeneous, important environments will remain undersampled. A practical solution is to partition active learning by structural motif, handling bulk phases, interfaces, and dislocations in separate acquisition loops in order to maintain extrapolation-grade accuracy and ensure comprehensive coverage of every relevant atomic environment. 

The clustering-enhanced local D-optimality scheme proposed in this study reduces the impact of structural heterogeneity by evaluating uncertainty within clusters of geometrically similar environments. This partitioned analysis maintains the accuracy of MLIPs and supports their transferability across defect-rich configurational landscapes. The approach is particularly helpful when expanding an existing database that already contains several defect classes. Unsupervised algorithms such as k-means or BIRCH\cite{Qi2024} can be used to divide the dataset into structurally coherent clusters before active learning or DIRECT sampling is applied. A comparable cluster-wise strategy could also be adopted for ensemble-based acquisition by assigning separate uncertainty thresholds to each subset; however, training many independent models would raise the computational cost substantially.

When using the original D-optimality method, it is important to note that the MaxVol algorithm focuses only on the most exotic atomic environments and therefore considers only the outer boundary of the dataset when constructing the active set. The major advantage is speed in the evaluation of the extrapolation grades, even for very large structures containing million atoms, but the drawback is reduced accuracy. In practice, the extrapolation grades calculated on heterogeneous datasets are often underestimated. Even with clustering-enhanced local D-optimality, capturing the fine details of every motif is difficult unless the acquisition step is carefully designed like Hodapp et al.\cite{Hodapp2020}. As a result, D-optimality-based active learning that starts from a global dataset tends to add very exotic structures or LAEs, which primarily guarantees the numerical stability of MD but may fall short of reproducing specific properties, such as dislocation migration or grain-boundary phase transitions, with DFT-level fidelity. A complementary route is provided by the QUESTS framework of Schwalbe-Koda et al.\cite{schwalbe2025model}, which measures the information-entropy increment that a candidate environment would contribute to a kernel-density estimate of the training distribution. Because this metric is model-free and depends only on geometric descriptors, it remains sensitive to rare motifs even in strongly heterogeneous datasets and can flag genuinely novel environments before any potential is fitted. Future studies could explore incorporating more expressive descriptors, such as SOAP\cite{SOAP-2013} or the message passing MACE representation\cite{batatia2022mace, Batatia2023, batatia2025design}, to better handle multi-element systems.


In summary, we have advanced the theoretical foundations of UQ for MLIPs within the ACE framework and delivered practical improvements that reinforce active‑learning workflows. By integrating high‑fidelity base models with both configuration‑ and atom‑resolved diagnostics, we enhance ensemble learning and D‑optimality’s capacity to detect truly novel atomic environments. We further expose the key failure modes that arise in heterogeneous configuration spaces and introduce a clustering‑enhanced local D‑optimality criterion that restores reliable uncertainty estimates across diverse datasets. These developments are essential for robust adaptive sampling and active learning, underpinning the efficient and confident development of MLIPs.

\section*{METHODS}\label{methods}
\subsection*{Machine learning interatomic potentials}

In this study, we use the ACE framework for UQ analysis for three main reasons. First, ACE\cite{Drautz2019, drautz2020, Lysogorskiy2021, Bochkarev2022, Lysogorskiy2023} provides a general, mathematically complete formalism\cite{Dusson2022} that can be extended to other descriptors such as Spectral Neighbor Analysis Potential (SNAP)\cite{Thompson2015} and MTP\cite{Shapeev2016, Novikov2021, Podryabinkin2023}. Second, it strikes an optimal balance between accuracy and computational efficiency\cite{Lysogorskiy2021}. Third, ACE’s built-in support for extrapolation-grade evaluation in both ASE\cite{HjorthLarsen2017TheAtoms} and LAMMPS\cite{Thompson2022} makes it straightforward to apply from small clusters up to million-atom configurations. To keep our analysis focused, we consider only linear ACE models. We explore six models of increasing complexity, ranging from 15 to 945 functions, covering a corresponding span of training accuracies. Throughout fitting, we fix the force-weighting parameter $\kappa$ at 0.01 and cap the number of training steps at 2,000.  The \textit{pacemaker} package manages ACE training\cite{Lysogorskiy2021,Bochkarev2022}.

\subsection*{Uncertainty quantification}
\subsubsection*{Ensemble learning}\label{ensemblelearning}

Following the Ref.\cite{Lysogorskiy2023}, we compute the maximum deviations of configurational energies and atomic forces, which serve as quantitative measures of uncertainty for each atom ($U_{F,\mathrm{atom}}$) and the whole configuration ($U_{E,\mathrm{cfg}}$ and $U_{F,\mathrm{cfg}}$), respectively, formulated as:
\begin{equation}
U_{E,\mathrm{cfg}} = \max_k\abs{E_j^k-\ave{E_j}}\,,
\label{eq:UQ-e-cfg}
\end{equation}
\begin{equation}
U_{F,\mathrm{atom}} = \max_k\abs{\mathbf{F}_i^k-\ave{\mathbf{F}_i}}\,,
\label{eq:UQ-f-atom}
\end{equation}
\begin{equation}
U_{F,\mathrm{cfg}} = \max_{i \in j}(\max_k\abs{\mathbf{F}_i^k-\ave{\mathbf{F}_i}})\,,
\label{eq:UQ-f-cfg}
\end{equation}
where $k = 1, . . . , K$ are the indices of the ACE models in the ensemble, $E_j^k$ is the energy predicted by model $k$ for the configuration $j$, and $\ave{E_j}$ is the ensemble average of the energy for the corresponding configuration. The force on atom $i$ in ensemble model $k$ is given by $\mathbf{F}_i^k$, while $\ave{\mathbf{F}_i}$ is the ensemble force average. Then, we compare the uncertainties $U_{F,\mathrm{atom}}$, $U_{F,\mathrm{cfg}}$ and $U_{E,\mathrm{cfg}}$, to their respective ground-truth errors, defined as:
\begin{equation}
e_{E,\mathrm{cfg}} = \abs{E_j^\mathrm{DFT}-\ave{E_j}}\,,
\label{eq:error-e-cfg}
\end{equation}
\begin{equation}
e_{F,\mathrm{atom}} = \abs{\mathbf{F}_i^\mathrm{DFT}-\ave{\mathbf{F}_i}}\,,
\label{eq:error-f-atom}
\end{equation}
\begin{equation}
e_{F,\mathrm{cfg}} = \max_{i \in j}\abs{\mathbf{F}_i^\mathrm{DFT}-\ave{\mathbf{F}_i}}.
\label{eq:error-f-cfg}
\end{equation}

In an active learning loop, new configurations or LAEs are selected when their uncertainties in predicted energy ($U_{E,\mathrm{cfg}}$) or force ($U_{F,\mathrm{cfg}}$ or $U_{F,\mathrm{atom}}$) exceed specified thresholds $\varepsilon_E$ or $\varepsilon_F$. Previous studies have typically relied on a force-based criterion, but the choice of $\varepsilon_F$ varies widely: hyperactive learning with linear ACE models often uses \num{0.2}-\qty{0.4}{eV/\angstrom}\cite{Erhard2024}, whereas active learning for MTPs in silicon-oxygen systems employs \num{1}-\qty{2}{eV/\angstrom}\cite{van2023hyperactive}. Lysogorskiy et al. proposes a consistent threshold for both energy and force:
\begin{equation}
\varepsilon = Q_3 + \num{1.5} \times \mathrm{IQR}\,,
\label{eq:threshold}
\end{equation}
where $Q_3$ is the third quartile of the training-error distribution of configurational energies or atomic forces and IQR = $Q_3$ – $Q_1$ is its interquartile range\cite{Lysogorskiy2023}. In this work, we adopt Eq.\ref{eq:threshold} because it automatically adapts to each ACE model’s specific training-error characteristics.

Using Eq.\ref{eq:UQ-e-cfg}, we establish the configuration-based energy criterion (CBE), noting that an atom-level energy criterion is physically meaningless since energy cannot be properly partitioned at the atomic scale\cite{Lysogorskiy2023}. Similarly, we derive two force-based uncertainty metrics from Eqs.\ref{eq:UQ-f-atom} and \ref{eq:UQ-f-cfg}: the atom-based force criterion (ABF) and the configuration-based force criterion (CBF). For small systems, we employ both CBE and CBF to detect novel configurations, while ABF serves as the primary metric for identifying new LAEs in large-scale simulations. A configuration is identified as novel if its energy uncertainty exceeds the threshold ($U_{E,\mathrm{cfg}} > \varepsilon_E$) or its force uncertainty surpasses the critical value ($U_{F,\mathrm{cfg}} > \varepsilon_{F,\mathrm{cfg}}$), while an atom is flagged as new when its local force uncertainty exceeds the threshold ($U_{F,\mathrm{atom}} > \varepsilon_{F,\mathrm{atom}}$).

\subsubsection*{D-optimality}

The \textit{pacemaker} package is used to construct the active set, and evaluates extrapolation grades ($\gamma$)\cite{Lysogorskiy2021, Bochkarev2022}. It should be noted that during active-set construction, \textit{pacemaker} removes outliers by discarding atoms whose forces exceed $\varepsilon = Q_3 + \num{1.5} \times \mathrm{IQR}$. By contrast, the standard MTP workflow retains every atom when computing $\gamma$\cite{Novikov2021, Podryabinkin2023}. This preliminary outlier filtering reduces D-optimality’s sensitivity to extreme configurations, providing a clear advantage over the conventional MTP approach (details are discussed in Section \nameref{data-hetero}).

Within standard active learning frameworks, two complementary extrapolation grades are typically employed: $\gamma_\mathrm{atom}$ (atom-level grade) and $\gamma_\mathrm{cfg}$ (configuration-level grade, defined as max $\gamma_\mathrm{atom}$ of each configuration). Conventional protocols flag configurations or atoms with $\gamma > 1$ as extrapolative, triggering the MaxVol algorithm to select determinant-optimizing environments for subsequent DFT calculations. These selected structures would then be added to the training set with active set updates, completing one learning cycle\cite{Lysogorskiy2023}. Although our present work focuses specifically on UQ methodology, this canonical active learning procedure provides important context for evaluating the performance of detection metrics and their implications for active learning efficiency.

\subsection*{Simulation and visualization}

We utilize the Vienna Ab initio Simulation Package (VASP) to perform first-principles calculations of all new configurations\cite{Kresse1996}. A gradient-corrected functional in the Perdew-Burke-Ernzerhof (PBE) form is used to describe the exchange and correlation interactions\cite{Perdew1996}. Electron-ion interactions are treated within the projector-augmented-wave (PAW) method, using the standard PAW pseudopotentials provided by VASP\cite{blochl1994projector}. The energy convergence criterion is set to $10^{-6}$ eV for electronic self-consistency calculations. The plane-wave cutoff energy is chosen to be 520 eV. The KPOINTS are generated by VASPKIT\cite{wang2021vaspkit}, based on the Monkhorst-Pack scheme\cite{monkhorst1976special}, with a consistent density of $2\uppi\times0.03$ $\unit{\angstrom}^{-1}$. Additionally, LAMMPS is used for force calculations and atomic extrapolation grade ($\gamma_\mathrm{atom}$) for million-atom configurations. OVITO is employed for the visualization of the atomic structures\cite{Stukowski2010}.

\section*{Data availability}
All ACE models and DFT datasets are available at the GitHub repository: \url{https://github.com/ufsf/UQ-ACE}.

\section*{Code availability}
All simulations are executed using open-source software LAMMPS. The machine learning force field was trained and validated by the \textit{pacemaker} package\cite{Lysogorskiy2021, Bochkarev2022}.

\section*{Acknowledgments}
This work was supported by the Nederlandse Organisatie voor Wetenschappelijk Onderzoek (NWO; the Netherlands Organization for Scientific Research), Domain Science, for access to supercomputing facilities. We also acknowledge the use of the DelftBlue supercomputer provided by the Delft High Performance Computing Center (DHPC; https://www.tudelft.nl/dhpc).

\section*{Author Contributions}
F. S.: Writing – original draft, Writing – review \& editing, Validation, Methodology, Data curation, Conceptualization. Z. W: Writing – review \& editing, Data curation and Analysis. K. L.: Writing – review \& editing, Data curation and Analysis. W. G.: Writing – review \& editing. P. D.: Writing – review \& editing, Supervision, Methodology, Funding acquisition, Conceptualization.

\section*{Conflict of Interest}
The authors declare that they have no known competing financial interests or personal relationships that could have appeared to influence the work reported in this paper.

\clearpage 

%

%
%
%
%
%
%


\newpage
\section*{Figures}

\newpage

\newpage

\newpage

\newpage

\newpage

\newpage

\newpage

\bibliography{ref} 
\bibliographystyle{sciencemag}

\newpage
\begin{center}
\section*{Supplementary Materials for\\ \scititle}
\end{center}

\subsubsection*{This PDF file includes:}

\noindent Supplementary Figure S1 to S2

\clearpage
\setcounter{figure}{0}
\renewcommand\thefigure{S\arabic{figure}}

\setcounter{table}{0}
\renewcommand\thetable{S\arabic{table}}

\newpage
\begin{figure}[!ht]
    \centering
    \includegraphics[width=0.8\linewidth]{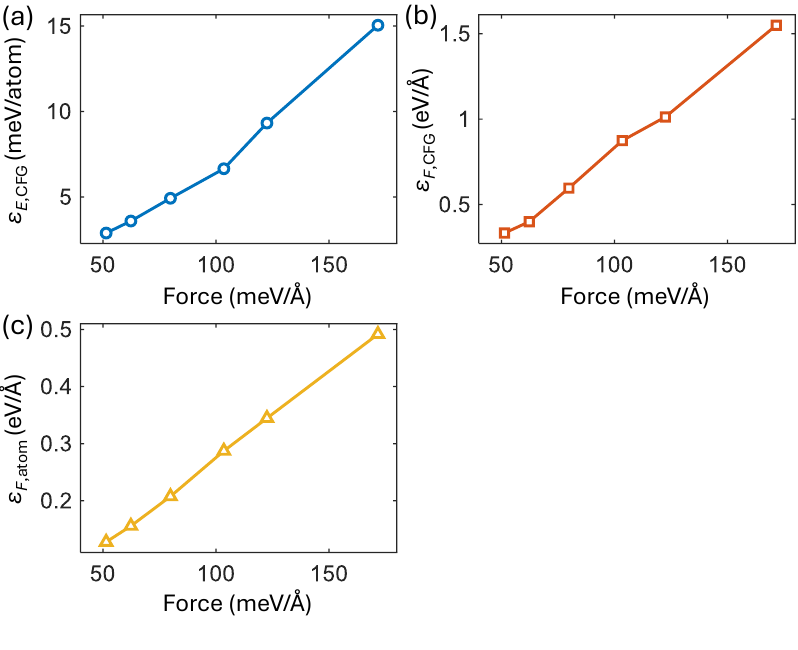}
    \caption{\textbf{Impact of model accuracy on different thresholds in ensemble learning}.}
    \label{fig:S1}
\end{figure}

\newpage
\begin{figure}[!ht]
    \centering
    \includegraphics[width=1\linewidth]{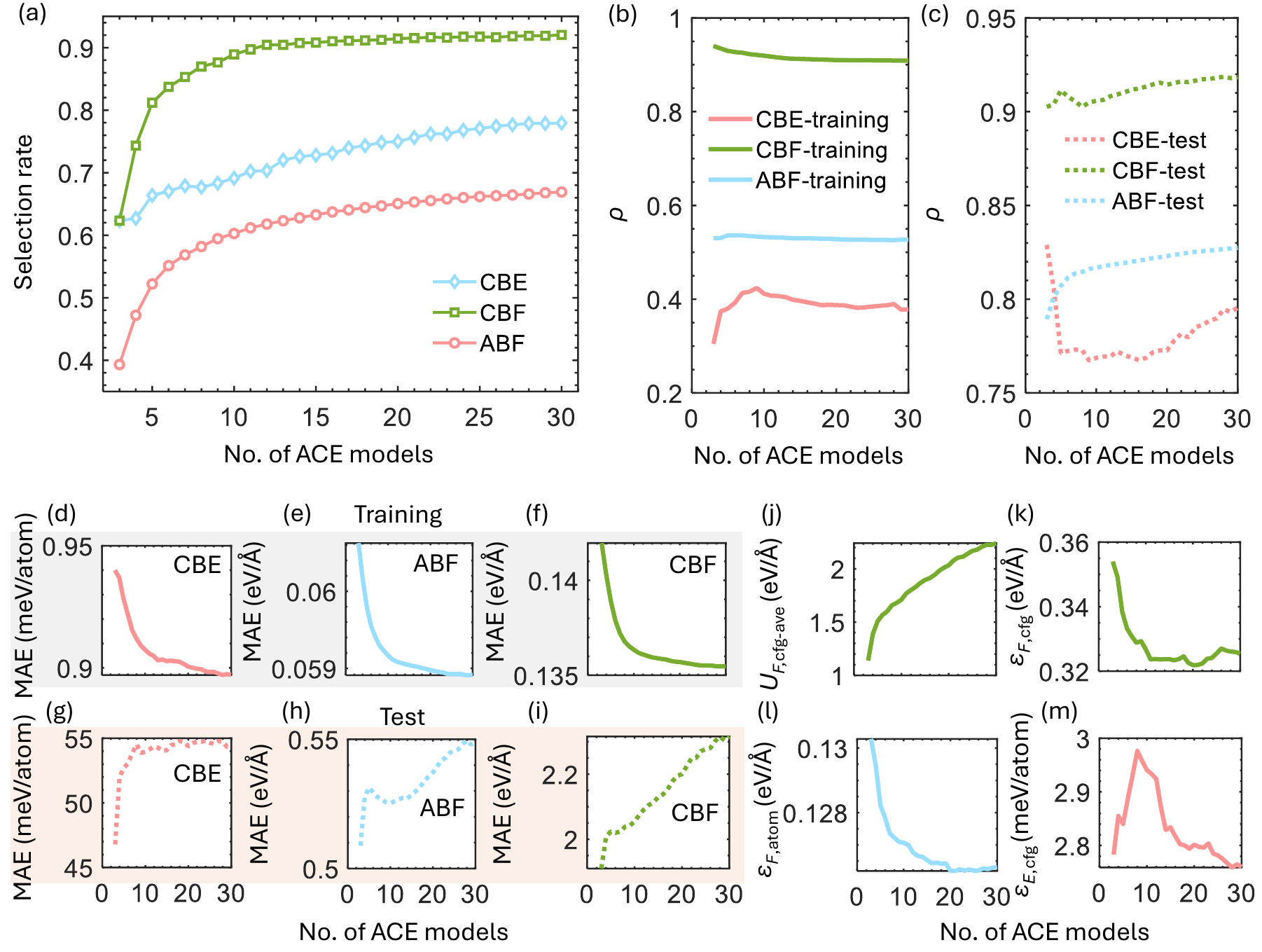}
    \caption{\textbf{Impact of ensemble size (number of Func-945 ACE potentials) on the reliability of atom and configuration selection.} \textbf{(a)} Detection rate as a function of ensemble size for all criteria. \textbf{(b, c)} Effect of ensemble size on the Spearman rank-order correlation $\rho$ between error and uncertainty for the training set \textbf{(b)} and test set \textbf{(c)}. \textbf{(d--i)} Effect of ensemble size on prediction errors for training and test datasets across three criteria. \textbf{(j)} Average uncertainty estimates for all test-set configurations using the CBF criterion, plotted as a function of ensemble size. \textbf{(k--m)} Effect of ensemble size on the thresholds used for detection of new configurations and LAEs.}
    \label{fig:S2}
\end{figure}

\end{document}